# Biocybersecurity: A Converging Threat as an Auxiliary to War

Lucas Potter[1], Orlando Ayala[2], Xavier-Lewis Palmer[1]

[1]Biomedical Engineering Institute, Old Dominion University, Norfolk, USA
[2]Department of Engineering Technology, Old Dominion University, Norfolk, USA
{Lpott005, Xpalm001}@odu.edu

**Abstract.** Biodefense is the discipline of ensuring biosecurity with respect to select groups of organisms and limiting their spread. This field has increasingly been challenged by novel threats from nature that have been weaponized such as SARS, Anthrax, and similar pathogens, but has emerged victorious through collaboration of national and world health groups. However, it may come under additional stress in the 21st century as the field intersects with the cyberworld -- a world where governments have already been struggling to keep up with cyber attacks from small to state-level actors as cyberthreats have been relied on to level the playing field in international disputes. Disruptions to military logistics and economies through cyberattacks have been able to be done at a mere fraction of economic and moral costs through conventional military means, making it an increasingly tempting means of disruption. In the field of biocybersecurity (BCS), the strengths within biotechnology and cybersecurity merge, along with many of their vulnerabilities, and this could spell increased trouble for biodefense, as novel threats can be synthesized and disseminated in ways that fuse the routes of attacks seen in biosecurity and cybersecurity. Herein, we offer an exploration of how threats in the domain of biocybersecurity may emerge through less foreseen routes as it might be an attractive auxiliary to conventional war. This is done through an analysis of potential payload and delivery methods to develop notional threat vectorizations. We conclude with several paradigms through which to view BCS-based threats.

## 1. Introduction:

The increasing intersections between the cyberworld and biotechnology have introduced novel threats, of which most can be defined as BCS Threats. These are threats at the nexus of biological, cyber, and physical apparatuses, allowing for malware to be distributed biologically or a pandemic-worthy agent via an autonomous drone guided over wi-fi networks. Teams across the spectrum of hacking, from the individual to the state level are looking at novel ways to attack and defend against cyberthreats from multiple angles, leading to deeper discussions on how. With the ability to modify aspects of nature such as photosynthesis to improve on food production or energy storage ever-improving the importance of nefarious purposes is worth consideration [1] Recently, many teams have brought these under-addressed possibilities to light, highlighting the many ways in which threats may emerge [2-3]. Through discussed vulnerabilities, exploits, and solutions differ, they point to a concerning issue: the morphing definition and basis of a payload that is transformed by this intersection. By payload, we refer to the functional aspect that causes damage to the systems that they are introduced to. Within, we discuss types of payloads and explore what could be used as the next delivery mechanism for potential future payloads, and ways of framing the use of bioweapons in hostile contexts.

## 2. Payloads:

A popularly used tool by penetration testers called Metasploit lists three types of payloads for the description of threat delivery that we can use here, singles, stagers, and stages [4]. Singles are self-contained; stagers link targets of interest and their attacker(s), and stages equip stagers. Physical examples of this, in order, can be direct exposures to a pathogen like an injection or sneeze (single), poorly developed applications that unnecessarily group people (stagers), and poorly ventilated buildings, and poorly planned infrastructures that amplify or provide the threats (stages). Adapting cybersecurity frameworks can give us a useful perspective on understanding intersectional threats.

The conventional payloads of biological warfare are bacterial and viral loads. These loads are typically extant microbes that could then be genetically modified to either incapacitate in the case of tularemia or affect a large number of fatalities, (for instance, weaponized smallpox [5-6]. The issue when it comes to deploying these agents as an auxiliary to warfare is, in one domain, well documented. Conventional dispersal of Anthrax, for instance, dates as far back to 1916 though it was deemed largely ineffective at the time [7]. The signing of the Biological Weapons Convention has eliminated most uses of biological weapons from conventional warfare [8]. Yet, the idea of using agents such as this for groups not bound by this treaty has never been greater, for the reasons elaborated below, including the lowered cost and increased accessibility to genetic modification, greater accessibility to targeting techniques (see Delivery Systems, below), centralized supply modalities (see in Delivery Systems), and an increasing number mechanisms by which to deliver a hypothetical biological agent (see Threat Vectorization) [9].

**3. Delivery Systems:**

The question of each threat's damage comes down to the mode of the exposure of the payload to the intended target. Is it intermittent like gusts down a hallway between bathroom breaks? Is it stealthy like hiding on a pen? Is it apparent, but unavoidable, like an elevator to an otherwise inaccessible penthouse? Is it segmented in delivery like meal courses at a company cafeteria? These must be pondered in considering the means and mode. The many methods by which a biological agent can be delivered is an intriguing notion when compared to the conventional threat analysis. Yet, the fundamental difference from the perspective of conventional threat analysis to threat analysis of BCS is that the delivery system for most conventional threats is built-for-purpose. The system of a bow delivering an arrow, or firearm delivering a bullet, or even a missile delivering a warhead, must be intricately designed to precise specifications for the delivery to be made. The delivery systems for biological threats can be entirely different. As seen in the Covid-19 pandemic, sneezing, coughing, touching, and even talking transform benign surfaces into vectors [10-11].

Knowing this, the fundamental problem of delivering a biological payload with modifications to make it more robust then turns not to be what the physical method of delivery is but rather targeting. Specifically, the problem is targeting people most likely to either spread or suffer from such an agent. This targeting can be surgical, involving one or a relatively few people, or it can be indiscriminate and en masse, for causing disruptions and indirect damage, making defense and containment difficult. A middle ground could be to identify genetic markers and utilize existing delivery systems to maximize cases of exposure. This could be useful for countering a condition for which many are asymptomatic but can be fatal in others. Take for instance the following scenario: a terrorist organization has made a bacterium that is especially virulent in the presence of a diabetic comorbidity. Assuming their goal is to maximize the number of casualties before any biosecurity groups can be alarmed, how might they best deliver their payload? Some ideas come to mind such as finding diabetic treatment centers and infecting shared surfaces-- doorknobs or waiting rooms. However, assume that they possess few resources to disseminate this agent and they want to minimize both their means of being tracked. This would require the minimal exposure of their members in spreading this contaminant. In this scenario, the possibilities become more limited, and they would need to find someone else to deliver their payload.

In this case, the centralization of shipping and logistics under the umbrella of electronic distributors becomes interesting. It is no secret that online retailers use information about their customers for marketing purposes [12-13]. If one makes an account on the site of an online retailer and browses items specific to people with diabetes (for instance test strips) they will quickly be shown hundreds of other items that people with diabetes may purchase. Here, assume there is an item of clothing that is exceptionally popular with people with diabetes. The factory that supplies this item, which is likely not subject to biosafety regulations or familiar with the threat then becomes a perfect target. The distribution and targeting of the payloads become simple questions of infiltrating the supply chain. Below are examples of everyday objects listed in a threat vectorization, combining which populations could match with which items, and so be targeted with a biological agent.

**4. Threat Vectorization:**

It is useful to examine how various objects in our present society could be used. In figure 1, we list some everyday objects to demonstrate how they could unknowingly be abused within a BCS context. These vectors are to be convenient delivery systems that fall in-between current threats. Blood/organ transplants and foodstuffs are excluded as they are subject to a battery of tests and routinely monitored for biological agents.

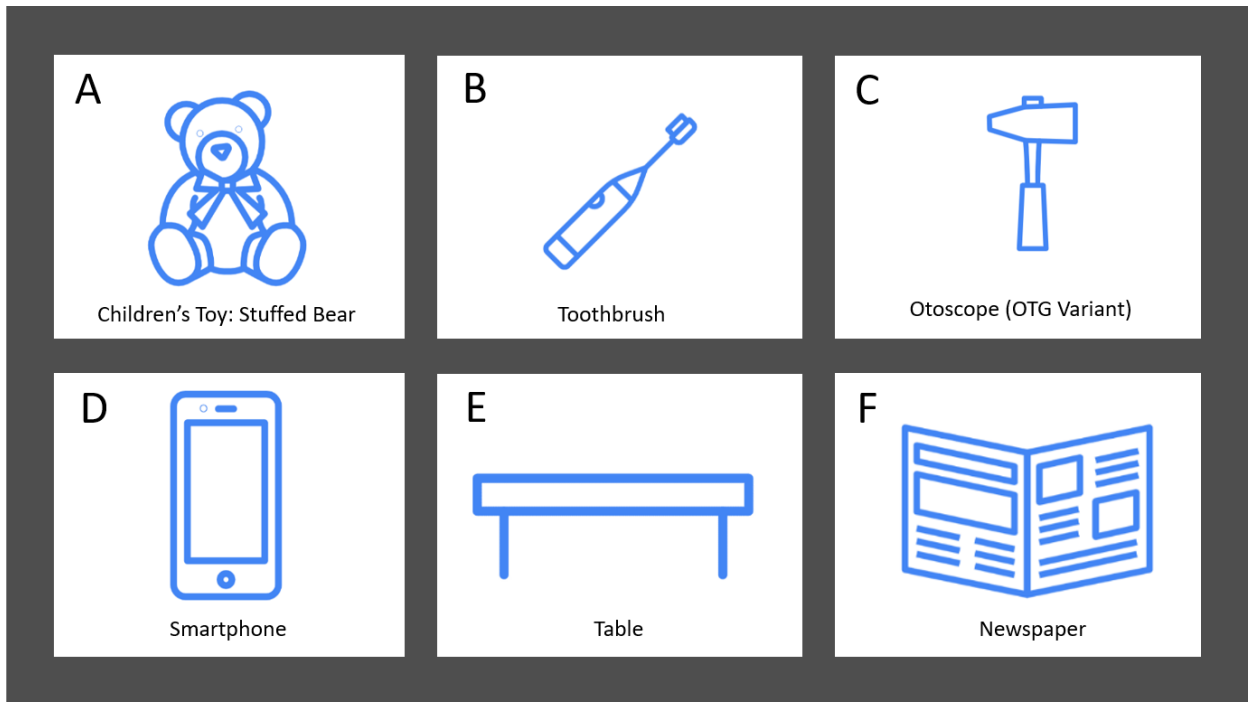

Figure 1: Common Items that could be solid infection routes or at worse, means of attacks.

A: Tufts of "fur" can trap dirt, dust, and host organisms at different scales. This can be amplified when they are wet or otherwise accumulate sweat and dead skin cells from kids who fail to frequently wash them.
Long amounts of skin to skin contact is possible, and this can be especially dangerous for immuno-compromised (specifically young) individuals. If these are made into smart-toys, it is important that sensors embedded and data transfer capacities are scrutinized to eliminate unnecessary leaks regarding health information of users, especially children. B: Toothbrushes are often wet and are rarely cleaned between uses, aside from rinsing. It is commonly guided for them to be simply disposed of after a few months. In terms of smart-products, these could be maliciously used to analyze personal data to identify optimal microbial threats in between uses, so caution is warranted with any sensors embedded and data collected. C: Otoscopes or similar medical peripherals: Devices like these are frequently used, and in times of fatigue, healthcare providers may fail to clean these, making for the possibility of pathogen transfer. Residual matter is commonly left on surfaces when healthcare providers fail to wash hands or change gloves, and in addition to failure to properly sterilize the non-disposable surfaces, and merely discarding the disposable parts can miss pathogens that have transferred at the disposable and non-disposable intersections. They sometimes interact with other devices, in a wired or wireless capacity. In wired capacity or slotted capacities, those devices also require cleaning, especially when connecting with other devices. Data hygiene should also be considered if there is data transfer and saving between smart devices, to minimize leaks of healthcare data. Concerning OTG (On the go) devices used in Telehealth, There aforementioned risk of contamination can be multiplied for the patient wherein insufficient training is given on the use and cleaning of said device as well as the risk of overly sharing or holding sensitive health information in terms of pictures of numerical health data. D: Smartphones are not often cleaned well and often taken to the bathroom, accumulating even more personal pathogens. it is common for these devices to touch and interact with a variety of surfaces such as laps, inner pocket linings, tables and desks, walls (when propped) and various holders, and more. Further, they hold many personal records and have sensors that can give data about the holder such as temperature or heart rate. Given the wide penetration of smartphones throughout the world, they are a ubiquitous and stealthy vector of compromise. Long amounts of contact with face and hands (which allows indirect contact with face) can improve transmission. E: Countertops and similar furnishings are exposed to the surfaces of many objects that are placed on it. Their vector potential can be exacerbated by ridge patterns or carpet-like material, which require more dedicated cleanings. In sum, surfaces may contain harmful agents due to cross-contamination or poor sanitation, and pores in surfaces may present difficulties for cleaning. Within BCS, designed agents can be spread and hidden on such surfaces as a means of attack or means of hiding data You can't slip a paper into a micro-sized pore, but you can slip bacteria with the contents of that paper in DNA within that same pore. F: Newspapers and similar magazines can not only host pathogens on their surfaces, but they can also disrupt airflow as their pages are turned due to their large surface area. When the user is alone, this is not problematic, but they need to be aware of surfaces that they have been exposed to such as when they transport these. In public, users need to be aware not to overly fan, to avoid fanning aerosols unknowingly across the public. More importantly, newspapers are never sanitized after leaving a central location and can interact with a wide number of people, resulting in the subtle, mass transfer of pathogens.

Above the idea of using modern supply chains and linking biological agents with cyber systems has been stated. Now, an elaboration of how these systems can be used is in order: BCS as siege warfare, BCS, and compromised telehealth, BCS in surveillance warfare.

**5.BCS as siege warfare:**

In popular culture, an act of bioterrorism is frequently thought of as a dagger in the back - something, one barely knew was there before it was utilized. It is time to add another way of thinking about biological threats. Ironically, biological warfare could see a return (especially if the current virus that has captured the attention of the world, Covid-19, could be weaponized) to its original form in the idea of siege warfare. Where once defenders flung infected masses over castle walls or by alternative methods, now the thought of using a biological contaminant to cause nations to shut down their borders, to cease internal traffic, and to occupy the skilled defenders has come full circle [14-15]. The notion of a biological agent overwhelming healthcare facilities and followed by infrastructural damage enhancing casualties or risk, has shown its effectiveness even as an accidental occurrence [16].

This becomes an even more important idea as more targeted methods of biological warfare become more available. Research into microfluidics, while nominally concerning itself with diagnosis, could be turned towards developing more virulent strains of biological warfare. Take the following case as an example of how the notional concepts of siege warfare could be used for a biological threat. If a group was able to effectively modify a biological threat to quarantine a population, they could then use targeted biological data (for instance a microfluidics unit embedded in an electric toothbrush) to release secondary or alternate biological agents with a greater affinity for the quarantined population to incite panic via a potpourri of specific biological agents. Variations of this could expand to different means of warfare which continue to occur with biology as a common link. We may see novel microbes designed and delivered to contaminate work sites that need to remain sterile, to degrade energy or food sources.

For instance, if a power plant serving a quarantined population has a large number of people that would be susceptible to a specific disease then after the quarantine was in place the plant itself could be targeted with an alternate vector. This would be much more effective than either of these two actions separately. This concept can be used in parallel with crafted information warfare campaigns. We could see carefully designed information campaigns designed to expose us to or augment current threats. This can be done more efficiently through populations that lack high science literacy and possess significant distrust in scientific institutions. As noted by the philosopher Amitabha Palmer, anti-science propagandists can prey on trust and our reliance on others for information, which creates a ripe atmosphere for sowing discord with well-resourced campaigns [17]. Recent happenings within information warfare, concerning Covid-19, serve as a clear example of where this risk is and can be further elevated. Wherein, foreign bots are encouraging individuals within America to re-open their economy and perhaps remove restrictions, hampering efforts to conduct an extended lock-down [18].

This can lead to an increase in infections, hampering not only the health of the populace but that of the economy, which of course affects a nation's military. BCS creates an indirect means of attack that cannot be understated without consequence. That is to say that such warfare is already here, intended or not, and it is up to us to consider how we may address it. This might be thought of as the only payload that is controllable as programs for informational dispersal can be shut off and on as needed, but once the material is out of the hands of the user or suffers some narrative disruptions, conspiracies can persist [19]. A question that merits asking those that consider deploying any biocyber threat is if they can justify or live with the threat persisting and eventually finding its way back to the country of origin.

**7.BCS and compromised telehealth:**

The payload can also be thought of digitally in the vein of telehealth, wherein patients use their phones or other computing telecommunication devices to consult with health professionals. Major points of vulnerability lie in the hardware used, but also infections that could confound the telehealth technologies used. For example, compromised sensors, compromised software, or the combination thief could allow compromised healthcare readings and outcomes. Deep learning enhanced malware was demonstrated to be effective in tricking radiologists on reading health outcomes for patients receiving a CT-scans, but this was for larger machines [20]. Who is to say that such could not eventually occur on smartphones and OTG accessories like otoscopes and auxiliary thermometers as machine learning becomes easier to adapt? In terms of sexual health, danger could apply to personal devices that could collect reproductive information, which additionally produces questions of care for marginalized sexual communities.

Further down the line, within the intersection of synthetic biology and cybersecurity, it is not unimaginable that state-level actors could create designer viruses that attack the dermis or epithelial tissues, producing difficult to detect QR-code analogs seen in the systems of birds, which indicate sex, hunger, and health qualities [21-23]. These could have the capability of compromising software designed to address and identify what a visual sensor picks up, especially that of software that uses system cameras similar to that of QR code scanning applications. One fix is through saving some aspects of care for in-person treatment, but a complication that may arise can also exist through exploiting vulnerabilities in means of indirect treatment through robots, as trailed in select cases [24-25]. The threat of in-person and telehealth sabotage remains and may grow in delivered medical goods, especially for future innovations in the development of new electronic prosthetics, the restoration of function in the form of electronic skin that can transmit sensation, and the beginning of research that may transmit taste [26-28]. We may see novel, but general attacks for the populace that eventually reach healthcare works with their prosthesis, confounding their ability to adequately assess a patient. One attack could be the equivalent of a digital Novocain or reflexive jerk, either of which could harm a patient, indirectly, through the provider's inability to feel for the use of their instrument on the patient, or directly through physical damage caused by involuntary movements. Herein is a scenario in which patients and/or healthcare workers are harmed and resources are wasted. Thus, concerns about telehealth at the intersection of BCS cannot be written off.

**8.BCS as surveillance warfare:**

The earlier mention of virus enabled skin modification addresses the intersection of enhanced surveillance. Through propagation of computationally optimized viruses that label individuals, such technology could enhance the detection and hunt for targets that might otherwise escape. One can imagine patterns caused on one's skin created by a virus that is difficult to see human eyes but is instead easily viewable and readable by special cameras. This code can morph with and report changes in one's metabolism, detailing drug use, people contacted, or beyond, and be generated on a commonly exposed area of skin. These might one day be transmittable, like a sexually transmitted disease. We illustrate the notion of reproducing tattoos in figure 2 below.

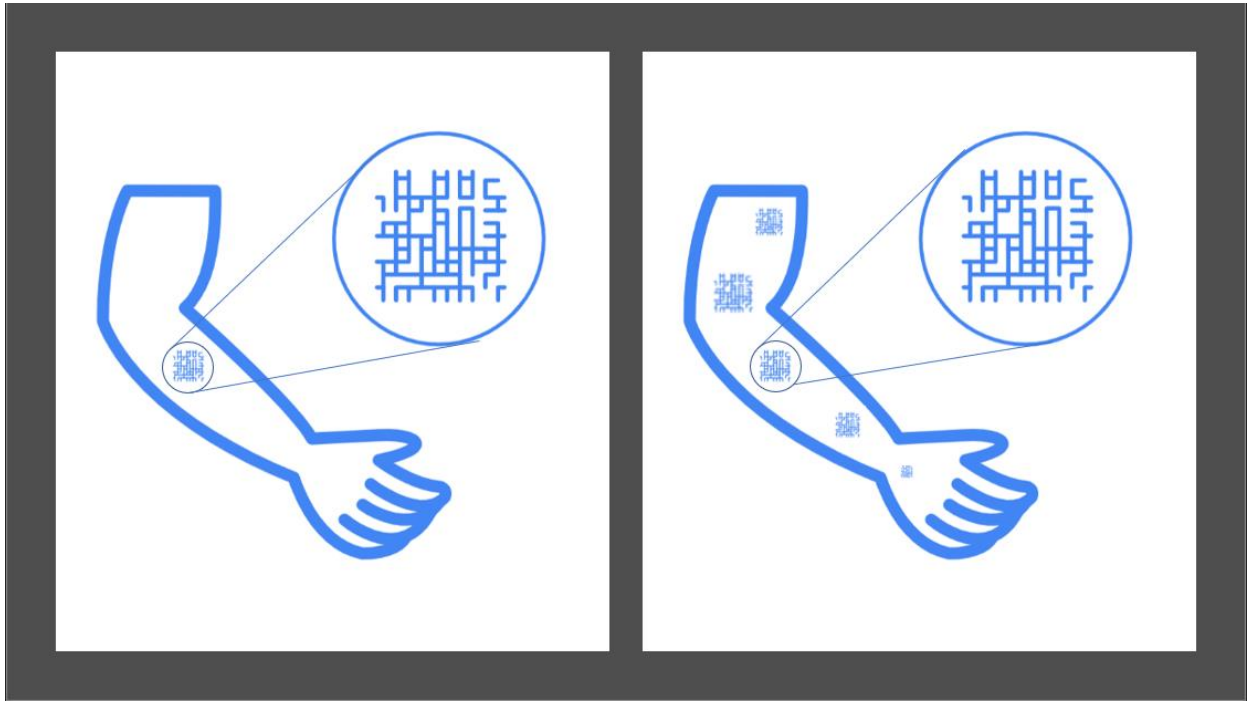

Figure 2: A picture of an imagined digital and functional tattoo on the left, along with the concept of multiple reproducing tattoos on an individual on the right.

Pictures that capture exposed skin with such code would allow for their surveillance as an individual is eventually to end up a mode of digitally accessible recording through social media. Photos can be geotagged, but this tag can be removed. However, subtle patterns in the skin of an individual would be harder to remove and plan for its removal in the same way as geotagging is watched for due to the lack of knowledge that one is aware that someone in their picture is tagged in such a surreptitious fashion. Often when an individual takes a selfie or picture within metropolitan areas, people in the background may be captured. Early on, this was not as much of a problem, owing to no or just low-resolution photography being available with the introduction of early cell phones. However, smartphone camera technology has been improving, allowing for the equivalent of a worldwide panopticon, especially given that cell phones of improving caliber are increasingly penetrating developing markets. One could be tracked in everyday photo documentation on social media without even knocking about it, around the world, due to accidental photobombing.

In military applications, a target of interest could have a device allowing tracking on a level with livestock as with RFIDs, QR codes, and or similar IoT technologies, but in a more, undetectable manner that has fused with your biology [29-30]. In this respect, resources allocated to track could be severely reduced, without sacrificing quality. Careful checks would be needed to curtail abuses such as ethics and security panels, improved public science literacy, and advocates sponsoring technology that counters the unauthorized and nonconsensual use of such. Groups working on technology to counter such is a given, and much like with the Cold War this could be one front of the bio-surveillance arms race. Further, emergent problems will need close inspections, as leaks and crosstalk occur, due to the increase in bio-insecurities produced, from this and the generation of information itself [31]. In terms of products to come out of this domain, a helpful means of public reassurance can come out of IoT capacity labeling as demonstrated by Carnegie Mellon's Cylab; with such products can be labeled for their surveillance capacity [31]. Disagreement is had over the nature to which bio-enabled surveillance and digitalization of biological aspects may be had, but what's undeniable is that institutions will need to reexamine the insecurities increased through the data generation and through laxes in best practices [31,33]. Actors, no matter their level, must tread lightly in this domain.

**9. General Cybersecurity Considerations:**

Cybersecurity as a means of conducting auxiliary actions to conventional warfare is a field with impressive depth and breadth- there are entire journals dedicated to cyber warfare such as the Cyber Defense Review. The exact protocols by which cybersecurity can be used adjacent to conventional military operations is too large of an issue to be constrained in this single article, though its importance to future defense operations is not to be understated. Simply detailing the different vectors by which one may obtain confidential information like passcodes could fill an entire text. This is an exercise for the reader, who can search the key phrase "Cybersecurity Textbooks" to find a plethora of writings. Searching "Biocybersecurity Textbook" will (at time of publication) likely return no or very few results.

However, it should be noted that a similar analysis of what is done here (particularly in figure 1) could be done for the cybersecurity domain. The authors actively encourage such research and would be curious to note what similarities, differences, analogies, and the possible new threats that could emerge from such an analysis.

**10. Potential Solutions:**

In the world of BCS, and the greater universe of threat analysis, there are rarely simple convenient solutions. Engineering rarely gives a clear-cut solution to any given problem, and instead leaves one with several piecemeal parts that solve added parts of the problem until the issue itself is negligible. An unsatisfying solution to threats mounted against the bio-cyber nexus is simply to end the cyber domain. But ask one of your colleagues if they could do their job without access to networked computers and you will find this solution untenable (and you would, likely and rightly, be mocked).

The threat of BCS is here to stay. Conventional solutions are the same ones we use for agricultural security. Regular monitoring, dedicated personnel, and robust legislation to protect the world's citizens from threats that they cannot control. One never truly considers that in the developed world, there are legions of people and libraries of regulations that control how their edible goods are gathered, processed, treated, and transported [34]. The primary issue is identifying which items should be compelled to be inspected. This would require a constant surveillance bioweapons program focused on longevity and robustness of bioweapons, constantly attempting to find the long-term viability of bioweapons on goods given the contemporary logistics network conditions.

**11. Conclusion:**

To conclude, there is a need to build upon our imagination as to what threat will emerge within the growing field of cybersecurity. Threats can emerge through the most mundane encounters and can unintentionally be delivered or be directed to. The question that remains is what each country will be doing to predict these routes, prevent, and counter them. Failure of a country to produce a reliable answer may spell their decline.

**12. Acknowledgements:**

We thank Google Autodraw for free image support. No external funding sources were used.